# Multi-channel, tunable quantum photonic devices on a fiber-integrated platform

*Woong Bae Jeon[1,†], Dong Hyun Park[1,†], Jong Sung Moon[1,2], Kyu-Young Kim[1], Mohamed Benyoucef[3], and Je-Hyung Kim[1,\*]*

[1]Department of Physics, Ulsan National Institute of Science and Technology (UNIST), Ulsan 44919, Republic of Korea

[2]Quantum Technology Research Division, Electronics and Telecommunications Research Institute (ETRI), Daejeon, 34129, Republic of Korea

[3]Institute of Physics, INA, University of Kassel, Heinrich-Plett-Str. 40, Kassel, 34132, Germany

*Corresponding Author: jehyungkim@unist.ac.kr,

†W. B. Jeon and D. H. Park contributed equally

**ABSTRACT**

Scalable, reliable quantum light sources are essential for increasing quantum channel capacity and advancing quantum protocols based on photonic qubits. Although recent developments in solid-state quantum emitters have enabled the generation of single photons with high performance, the scalable integration of multiple quantum light sources onto practical optical platforms remains a challenging task. Here, we present a breakthrough in achieving a multiple, tunable array of quantum photonic devices. The selective integration of multiple quantum dot devices onto a V-groove fiber platform features scalability, tunability, high yield, and high single-photon coupling efficiency. Therefore, our fiber-integrated quantum platform realizes a scalable and reliable single-photon array within a compact fiber chip at telecom wavelengths.

Recent advances in solid-state quantum emitters have enabled the efficient generation of single and entangled photons with high purity, indistinguishability, and fidelity.[1-3] For practical and reliable utilization of these quantum photonic resources, the heterogeneous integration of quantum emitters with low-loss optical fibers is essential in future quantum applications across optical fiber networks.[4] So far, fiber-integrated quantum emitters have been realized through several approaches, including a lensed fiber,[5] micro-optics,[6] and adiabatic coupling.[7,8] More practical plug-and-play type single-photon sources have been demonstrated by directly gluing a fiber on a quantum dot (QD) wafer[9-11] and also directly integrating a single QD device onto a fiber.[12,13]

While such fiber-coupled single-photon sources would be useful for a simple quantum key distribution scheme,[14] the most advanced photonic quantum applications, such as quantum neural networks[15] and quantum simulators[16,17], require multi-channel quantum light sources. Multiple single-photon sources are also useful for creating large-scale quantum entanglement, such as high-dimensional photonic graph states.[18] A spatial demultiplexing technique with temporally separated photons from a single QD has been introduced to realize a multiple array of single photons and has demonstrated multi-photon boson sampling.[16] Whereas the demultiplexing method efficiently generates indistinguishable multiple single photons from a single emitter, it sacrifices the generation rate with increasing channels. Alternatively, generating single photons from individual quantum emitters at each channel provides more versatile multiple single-photon arrays. In previous approaches, periodically patterned QD devices on a chip have been coupled with multi-channel fibers via precise alignment[9] or photonic wiring.[19] However, this chip-to-fiber integration lacks the efficiency, selectivity, and tunability of the coupled quantum emitters. In particular, solid-state quantum emitters suffer from large randomness in their positions and frequencies. Therefore, this simple chip-to-fiber array integration has a low yield of the integrated devices, and each channel generates single photons with random frequency, which poses significant limitations to being deployable sources for quantum applications.

Here, we demonstrate multi-channel and tunable single-photon sources on fiber-integrated platforms. We address previous challenges by developing a multi-functional fiber platform that selectively and efficiently integrates multiple QD devices. Furthermore, the integrated electrodes on the fiber platform show a potential for spectral tunability of single-photon emissions. Therefore, our integration, interfacing, and engineering techniques enable scalable quantum light sources with high yield, high efficiency, and controllability.

Scalable integration of multiple quantum emitters was conducted by transferring individual telecom-emitting InAs/InP QD devices onto optical fibers.[13] The QD sample used was grown by molecular beam epitaxy on an InP substrate[20,21] and shows a broad spectral distribution from 1,200 nm to 1,400 nm (Figure S1). A 1400 nm InGaAs sacrificial layer was initially deposited to fabricate air-suspended cavities. This was followed by the growth of a 150 nm thick InP layer, subsequent growth of InAs QDS, and finally, a 150 nm thick InP capping layer was grown. In order to realize fiber-integrated single-photon arrays, we first developed a multi-functional fiber-integrated platform. Figure 1(a) describes a schematic image of QD devices integrated on a multi-channel V-groove fiber array. QDs require cryogenic temperature to generate single photons with high purity and indistinguishability, whereas the integrated fibers lack thermal conductivity. To resolve this cooling issue in the fiber-integrated quantum devices, we coated the front surface of a V-groove silicon fiber block, including a standard SMF-28 fiber array, with a 50 nm thick gold film. Then, we etched the metal film at fiber cores to a depth of over 70 nm using a focused ion beam. Our QD membrane devices are designed to be slightly larger than the etched metal hole. Therefore, the deposited gold film effectively cools down the QD device, and the etched metal at the fiber core serves as an optical window for single-photon transmission from the integrated QDs. Then, the metal-coated Si fiber block was mounted on the metal mold connected to the cryostat. Therefore, the fiber-integrated QDs are thermally connected through a coated metal thin film, a Si block, and a cold finger of the cryostat and can be cooled for low-temperature operation (10 K). In addition, to apply electrical bias, we grounded the deposited gold film on a fiber facet and placed another electrode at a distance. Therefore, two electrodes apply an electric field to the QDs for spectral tuning.

Coupling single photons from QDs into a single-mode fiber demands optimal optical interfaces to improve light extraction and mode-matching conditions. A thin-membrane structure with a nanophotonic cavity design provides a suitable interface for stable device-to-fiber integration. For example, a well-known ring-based circular Bragg gratings structure[22,23] could generate a vertical Gaussian beam. However, the cavity mode and fiber guide mode show a mismatch in their mode size due to the large refractive index differences between the two materials, which limits the single-photon coupling efficiency from a QD to a fiber. Introducing a hybrid structure could enhance the coupling efficiency, but it requires a complicated reflector design.[24] Recently, we have developed hole-based circular Bragg grating (hole-CBG), which generates a highly directional Gaussian beam that matches the small numerical aperture of an optical fiber.[13,25] The major advantage of hole-CBG cavity comes from reduced refractive index contrast by azimuthally distributed hole arrays on a subwavelength scale.[26] This reduced refractive index contrast spreads out a spatial cavity mode widely, as shown in Figure 1(c), leading to a reduction in the divergence in momentum space according to the Fourier transform relationship. We optimized the hole diameter and axial and radial distance between holes and achieved 60 percent of far-field emission within 7 degrees (Figure 1(d)). As shown in Figure 1(b), from the simulation of a fiber-integrated hole-CBG device, including a partial reflection by the top electrode, we calculate a fiber coupling efficiency of 58.5% for an optimized electrode distance of about 800 nm. Together with the optimized far-field emission, we calculated a Purcell factor of 120 at a cavity mode having a calculated quality factor of 4,000. (See Supporting Information for the details on the cavity design and numerical simulation results).

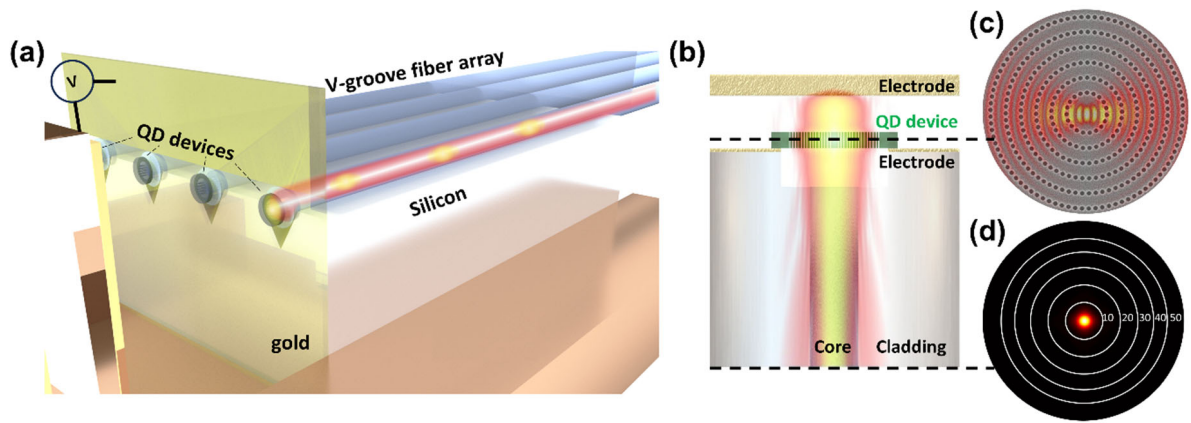

**Figure 1.** (a) Schematic image of a multi-channel, tunable single-photon array based on a fiber-integrated platform. (b) Detailed cross-sectional configuration of a fiber-integrated QD device with metal layers for low-temperature cooling and applying an electric field. A cross-sectional electric field profile is superimposed on a log scale. (c) SEM image of a fabricated InAs/InP QD cavity with a superimposed near-field cavity mode profile. (d) Calculated the far-field profile of the dipole-coupled cavity.

We fabricated the designed hole-CBG devices using electron beam lithography followed by dry and wet etching processes on an InAs/InP QD sample (Figure 1(c)). Figure 2(a) shows a transfer process of individual hole-CBG devices under an optical microscope. For picking up a single QD device, we used a polydimethylsiloxane (PDMS) microstamp with a radius of 20 μm and a height of 7 μm (inset in Figure 2(a)). We brought this cylindrical microstamp closer to each device, picked the device up, and integrated it on each channel of the V-groove optical fiber array. Figure 2(b) shows microscope images of a gold-coated V-groove fiber array with an etched fiber core before (top) and after (bottom) the device transfer. Figure 2(c) shows that the QD device is firmly integrated on a single optical fiber by the Van der Waals force.

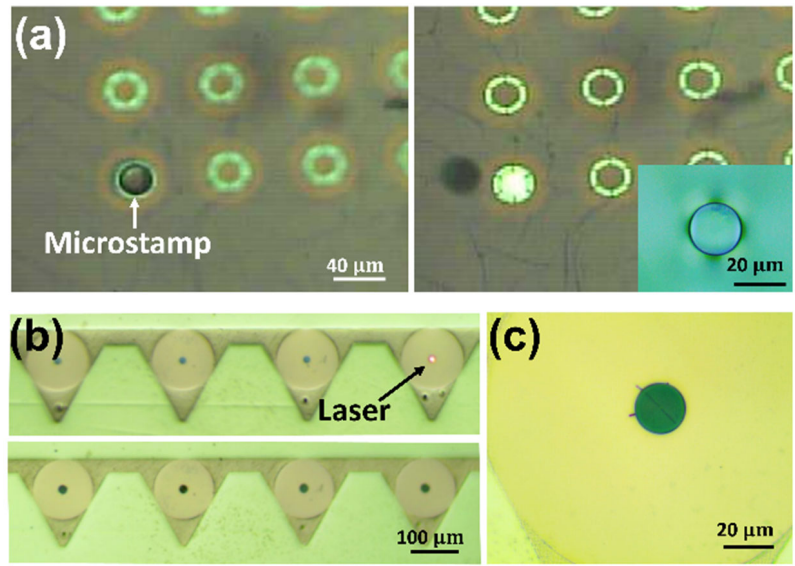

**Figure 2.** (a) Optical microscopy images captured during transfer of a QD device using a PDMS microstamp. The left (right) image shows the microstamp approaching (picking up) the QD device. The inset shows the magnified PDMS microstamp. (b) Microscopy images of a gold-coated V-groove fiber array before (top) and after (bottom) the transfer of QD devices. A 633 nm red laser passing through the etched fiber cores is clearly visible in the top image. The bottom image shows four QD devices integrated onto each fiber port. (c) A closer view of the fiber-coupled QD device.

We optically characterize single-photon emission from the fiber-integrated QD devices at a low temperature of 10 K. Figure 3(a) describes a schematic of an all-fiber connectorized single-photon system from a laser to a single QD and single-photon detectors. A fiber-coupled 785 nm laser was sent to the fiber-quantum channel by a 10:90 beamsplitter and excited the QD device. The photoluminescence (PL) signal from the QDs was coupled to a fiber and transmitted to a high-resolution spectrometer for spectrum measurements or a fiber-based spectral filter followed by a fiber-based Hanbury Brown and Twiss (HBT) interferometer for photon-correlation measurement. Therefore, our system operates on an all-fiber basis without any optical alignment. Figure 3(b) displays an optical image of the multi-channel single-photon array mounted on a cryostat. Electrical signals were also connected to our fiber platform to induce voltage.

Figure 3(c) shows the PL spectrum from the QD devices integrated on fiber Port 1 to 3. Each fiber channel shows cavity-enhanced single QD emissions at 1,220, 1,228, and 1,237 nm with spectral linewidths of 11, 23, and 12 GHz, respectively. To confirm the fiber-coupled QD device emits single photons, we performed an HBT experiment for the spectrally filtered single QD emission on Port 3. The photon correlation data in Figure 3(d) clearly show an antibunching signal with $g^{(2)}(0) = 0.06 \pm 0.02$ Together with single-photon purity, indistinguishability is also an important feature of single photons. We characterized the indistinguishable nature of single photons by performing fiber-based Hong-Ou-Mandel type two-photon interference (Figure S6 for the schematic image of the setup). Figure 3(e) shows the resulting photon correlation histogram ($g^{(2)}_{\parallel}(\tau)$ and $g^{(2)}_{\perp}(0)$ ) for co- and cross-polarized conditions by controlling fiber-polarizing optics. While the cross-polarized two photons show increased coincidence counts, the co-polarized two-photon case exhibits suppressed coincidence counts at a delay time of zero as a result of quantum interference between two identical photons. The measured visibility at delay time zero is around 0.66±0.02 with a coherence time of 151±7 ps (Figure 3(f)). Non-unit visibility could be attributed to nonzero $g^{(2)}(0)$ in HBT, imperfect polarization controls, and unbalanced intensities in two fiber input channels. In addition, the short coherence time of the QD compared to its spontaneous emission time of 700 ps limits the temporal window for achieving indistinguishable single photons. Lifetime-limited linewidth has been achieved in other QD platforms by utilizing (quasi) resonant excitation techniques and by reducing charge noises.[27,28]

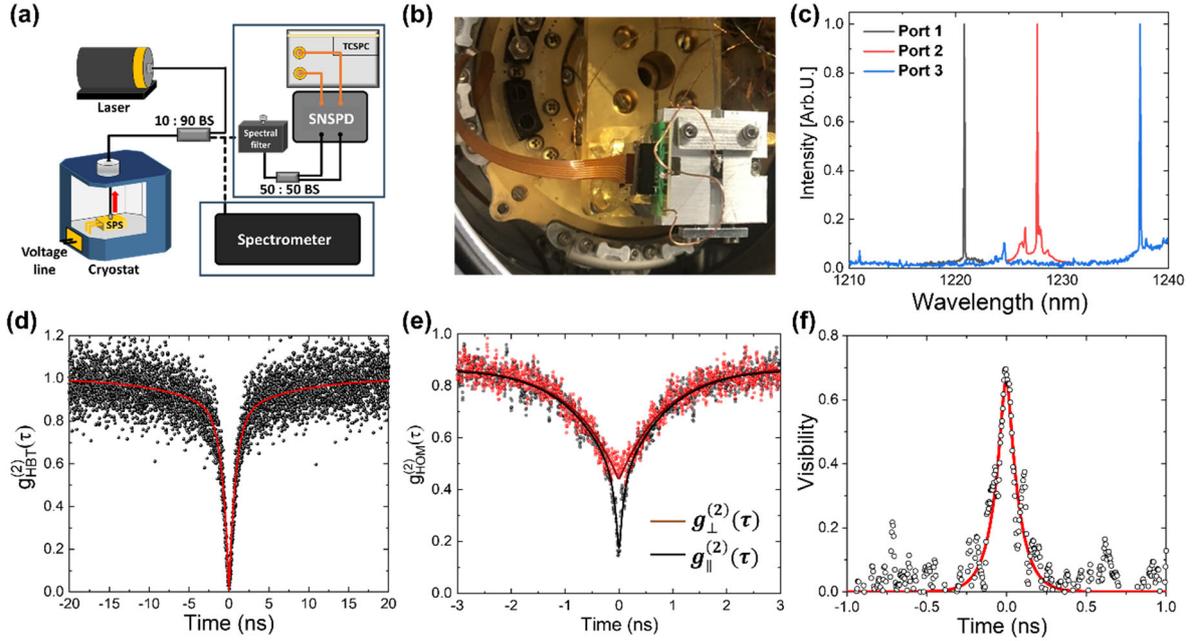

**Figure 3.** (a) Schematic of the all-fiber-connectorized optical setup for single-photon characterization. (b) Photo image of a multi-channel single-photon array mounted on a cryostat sample holder. (c) PL spectra of fiber-integrated QD devices on Port 1 to 3. (d) HBT histogram of a single QD emission from fiber Port 3. (e) Two-photon interference for co- and cross-polarized photons from fiber Port 3 using fiber-based Hong-Ou-Mandel interferometer. (f) Two-photon interference visibility from the data in (e). Red lines represent fitted curves.

The important advantage of our device-to-fiber integration is that it enables the pre-characterization of several QD devices from their original wafer and the selective integration of specific devices with desired wavelengths. From fabricated cavity devices on randomly distributed high-density QD ($\approx 10^9 \ cm^{-2}$) wafer, we could easily find cavity-coupled QDs near the low $Q$ (460) cavity mode. We quantified the yield (4%) of finding single QDs at a target wavelength within $\pm 1 \ nm$ spectral window (Figure S2). However, to bring the advantage of precharacterization, the device properties should remain consistent after the device transfer process. In particular, the direct transfer of QD devices onto a fiber facet having a refractive index of about 1.45 results in a spectral shift of the cavity mode. To avoid this change, we etched the fiber core by 70 nm using a focused ion beam, allowing the transferred cavity to remain air-suspended even after fiber integration. Additionally, the interconnected design of hole-CBGs offers more robustness against potential deformation by the heterogeneous integration compared to conventional ring-CBGs. Therefore, our approach enables pre-characterization and selective integration of multiple QD devices. To demonstrate this capability, we searched for the specific cavity-coupled QD device having a similar wavelength of around 1,237 nm to the QD device on fiber Port 3. Figure 4(a) compares the PL spectrum of the selected QD device before and after the fiber integration. The comparison spectrum in Figure 4(a) shows that the air-suspended integration with a microstamp preserves the cavity mode and QD frequency well before and after the transfer process, validating the spectral identity after fiber integration. We attribute the observed small wavelength shift (0.144 nm) of the QD peak to the small amount of induced tensile strain. In Figure 4(b), we display PL spectra from two separate fiber-integrated QD devices on Ports 3 and 4. Two fiber-integrated QD devices show nearly resonant emission wavelengths. From the Lorentzian fitting we calculate the spectral detuning of 0.024 nm. Such selective integration was impossible with previous chip-to-fiber integrations.[9,19] This aspect holds significant importance in achieving multiple, identical quantum emitters with high yield and establishing scalable quantum interference or entanglement.

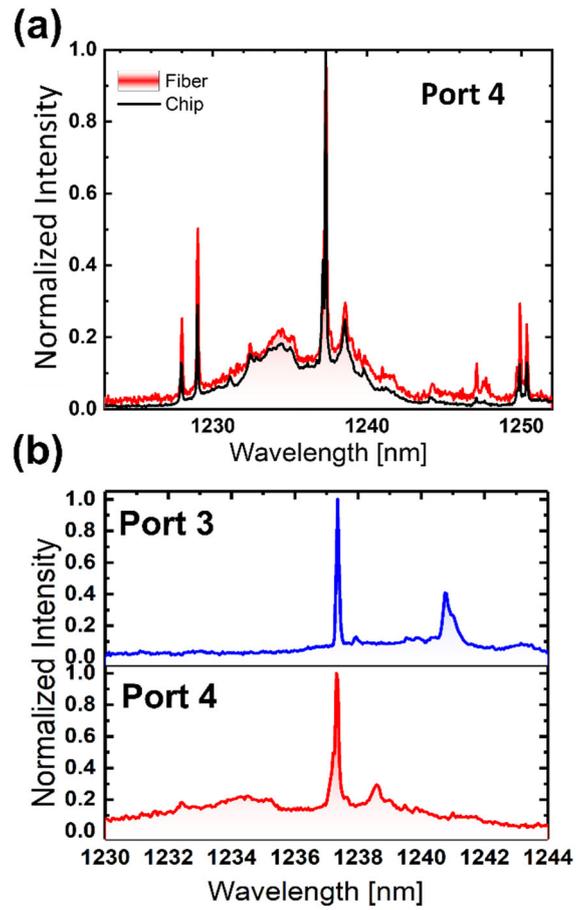

**Figure 4.** (a) Comparison of PL spectra of the same QD device measured before and after fiber integration on Port 4. (b) Comparison of PL spectra from two induvial QD devices integrated on fiber Ports 3 and 4.

Although pre-characterization and selective integration secure multiple quantum channels with target frequencies, additional fine frequency tunability will be desired to eliminate any remaining detuning from the target frequency. To accomplish frequency tunability on fiber-integrated, we apply an electric field through the fiber-integrated electrodes on the front and back of QD devices. The back electrode was fabricated on a fiber facet and directly contacted the QD devices at the edge of the cavity, while the front electrode was formed by a separated panel having spatial distance from the cavity. We manually controlled the distance between the electrodes with screws by monitoring the interference fringes depending on their distance. We estimate the distance between two electrodes to be about 7.7 μm. Increasing the applied voltage shifts the frequency of the QD device, as shown in Figure 5. At a voltage of 160 V, an electric field of up to 207.8 kV/cm is applied to the device. The results demonstrate the potential tunability of single-photon emission on the fiber-integrated platform, and the observed amount of frequency shift is similar to heterogeneously integrated QD devices on a photonic circuit platform.[29] However, even with the adoption of precharacterization and selective integration techniques, the current tuning range was relatively small compared to a typically required tuning range of around 1 meV. In this work, the far distance between manually controlled two electrodes leads to a small Stark shift compared to previously demonstrated widely tunable QDs that experience a significantly large electric field through a p-i-n structure.[28,30,31] Therefore, to extend the tuning range in our platform, a piezo stage could be adopted to precisely control the electrode distance, or more effectively, a similar p-i-n structure could be implemented. In addition, introducing quasi-resonant excitation could also increase a tuning range by reducing a field screening effect.[29]

In addition, our current platform has a simple configuration of two electrodes, which does not allow independent tuning of single-photon emitters on each channel. Developing multi-channel microelectrodes would enable independent frequency tuning. This capability can be particularly useful for scalable quantum applications requiring precise control over multiple emitters, such as multi-photon interference experiments[16] and increased channel capacity in quantum communications.[32]

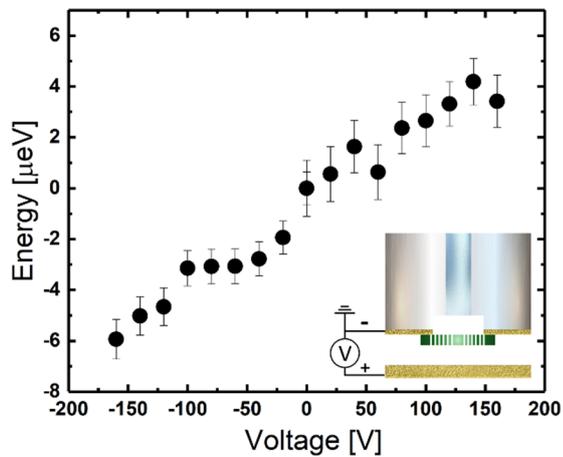

**Figure 5.** Frequency tuning of a single QD emission by fiber-integrated electrodes. The inset shows the configuration of an applied electric field through a QD device using two separate electrodes.

Finally, we quantify the coupling efficiency of fiber-integrated QD devices. We separately measured system efficiency of 35.6% (Supporting Information), including the transmission losses in fiber connectors and a spectral filter and the detection efficiency of single-photon detectors. Then we excited the QDs with an 80 MHz pulsed laser and calculated the single-photon coupling efficiency of 6.5%, 9%, 7%, and 6% at the end of the first fiber from the single QD on fiber channels from 1 to 4, respectively. The reason for lower coupling efficiency in experiments than the simulated value would be the small misalignment between the cavity center and QD devices and the uncontrolled position of QDs to the cavity center (Figure S5). This could be improved by introducing site-controlled QD growth[33] or cavity fabrication[3,34] techniques. Furthermore, optimizing the etching depth of the fiber could increase the coupling efficiency further.[35]

In summary, we have demonstrated fiber-integrated multi-channel, tunable single-photon sources. By combining an optimal nanophotonic cavity, device-to-fiber integration, and applying electric field techniques, the multi-functional fiber platform exhibits capabilities of low-temperature cooling, efficient single-photon transmission, selective integration, and frequency tunability, thereby enhancing the overall performance and adaptability of our fiber-integrated quantum photonic devices. This fiber-integrated platform directly interfaces all optical and electrical signals, eliminating the need for additional alignment processes. This offers a scalable and reliable quantum photonic resource for advanced applications. In particular, together with the matured fiber optics technologies and fiber-coupled multi-channel single-photon detectors, the development of fiber-integrated multi-channel quantum light sources can establish all-fiber-connected real-world quantum communications.[14,36,37]

**Notes**

Any additional relevant notes should be placed here.

**ACKNOWLEDGMENT**

This work is supported by the National Research Foundation of Korea (RS-2024-00438839, 2022R1A2C2003176, RS-2024-00442762), IITP (IITP-2024-2020-0-01606, RS-2023-00259676, RS-2023-00227854). M. Benyoucef acknowledges the support by the German Research Foundation - DFG (DeliCom, Heisenberg grant-BE 5778/4-1) and BMBF (QR.X). We thank R. Kaur, A. Kors for their assistance in the MBE growth process, J. P. Reithmaier for his discussion, and Dirk Albert for his technical assistance. The authors would like also to thank Prof. Edo Waks from the University of Maryland and C. J. K. Richardson from the Laboratory for Physical Sciences, University of Maryland, for providing initial testing QD samples.

# Supplementary Information

## 1. Characterization of quantum dots on a chip

### 1) PL spectrum of quantum dot ensemble in a bulk chip

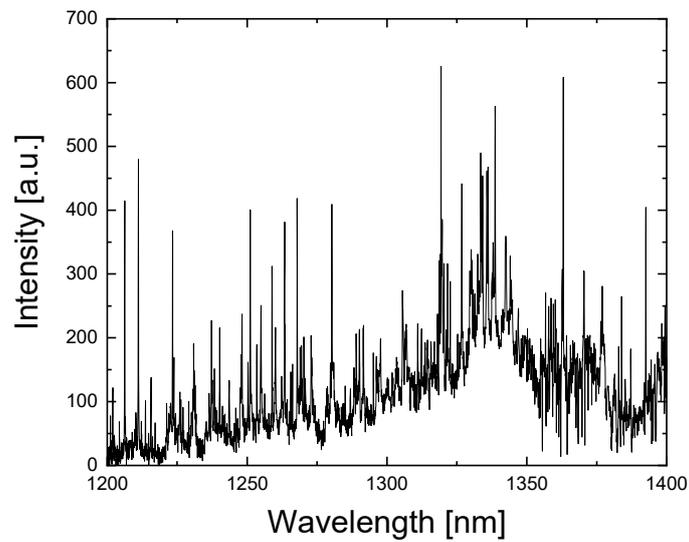

**Figure S1. Low temperature (4 K) PL spectrum of quantum dot ensemble in a bulk chip.** The emissions from the quantum dot ensemble are widely spread over 1200 nm to 1400 nm.

## 2) Precharacterized and postselected cavity-coupled quantum dots

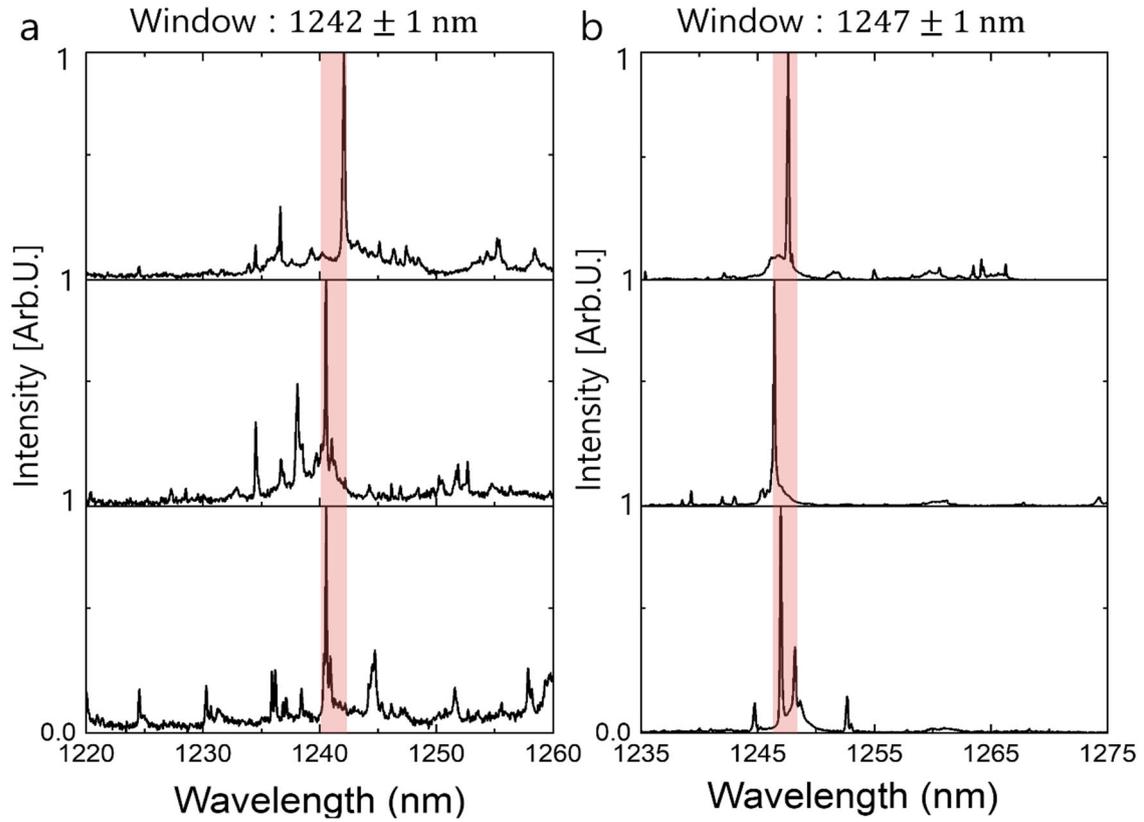

**Figure S2. PL spectrum of cavity-coupled quantum dots.** (a,b) Cavity-coupled quantum dots with cavity modes at 1242 nm (a) and 1247 nm (b). By statistically characterizing 60 different cavity devices, we calculate the yield of finding such quantum dots at target wavelengths within $\pm 1$ nm to 4 %.

## 2. Single-photon coupling efficiency of fiber-integrated quantum dots

To calculate the collection efficiency at the first fiber, we first measured the transmission efficiency of the system components. The schematic of the system is illustrated in Figure S3, and the table on the right summarizes the transmission efficiency of each component as well as the detection efficiency of a superconducting nanowire single photon detector (SNSPD). The system efficiency ($\eta_{sys}$) in total was determined to be 35.6% of our all-fiber-based system.

To measure a fiber-coupling coupling efficiency, we excited a quantum dot using a 785 nm pulsed laser with an 80 MHz repetition rate. At this repetition rate, the photon count rate ($\gamma$) is related to the efficiencies as follows;

$$80 \text{ MHz} * \eta_{int} * \eta_{coupling} * \eta_{sys} = \gamma \qquad (1)$$

, where $\eta_{int}, \eta_{col}$ and $\eta_{sys}$ represent the internal, coupling, and system efficiency, respectively. At low temperature, we assumed the internal efficiency ($\eta_{int}$) to be 1. From the calculation above with the measured photon count rate ($\gamma$) of 2.57 MHz, we estimated the fiber-coupling coupling to be 9% for the brightest fiber-coupled QD in Port 2.

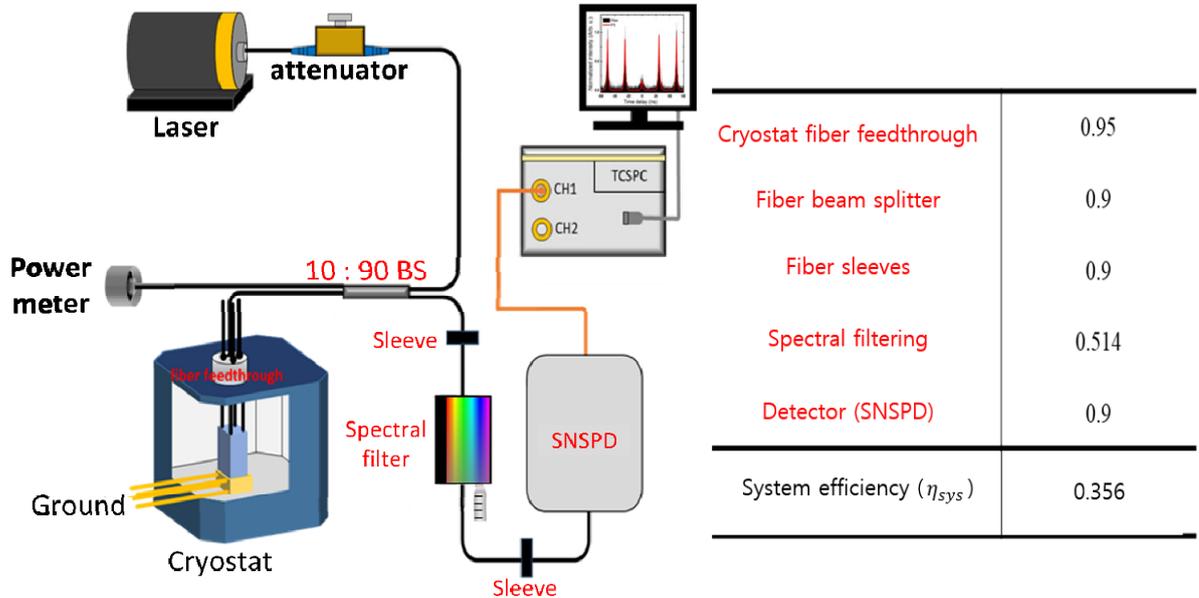

**Figure S3. Schematic image and efficiency of the all fiber-coupled system.**

## 3. Numerical simulation of a cavity mode

### (1) Optimization of a cavity mode for highly directional emission

We used finite-difference time-domain (FDTD) simulations to maximize the collection efficiency of a hole-circular Bragg grating (H-CBG). **Figure S4(a)** schematically illustrates the H-CBG structure and its adjustable parameters. We considered that the radius of the center disk (c), hole size (h), and axial period (a) are linearly related to the radial period (Λ). We initially set Λ = 467 nm and swept through the other parameters to maximize the collection efficiency, ultimately determining the optimal values as c = 618 nm, h = 135 nm, and a = 210 nm for an InP ($n = 3.2$) membrane (300 nm). Figure S4(b) compares the angular distribution of far-field emission from an in-plane dipole in bulk InP and an H-CBG cavity. **Figure 4(c)** shows the increased Purcell factor of the designed H-CBG near the cavity mode at 1249 nm. The widely enhanced collection efficiency is also plotted as a function of a wavelength, considering collecting numerical apertures of 0.7 and 0.12.

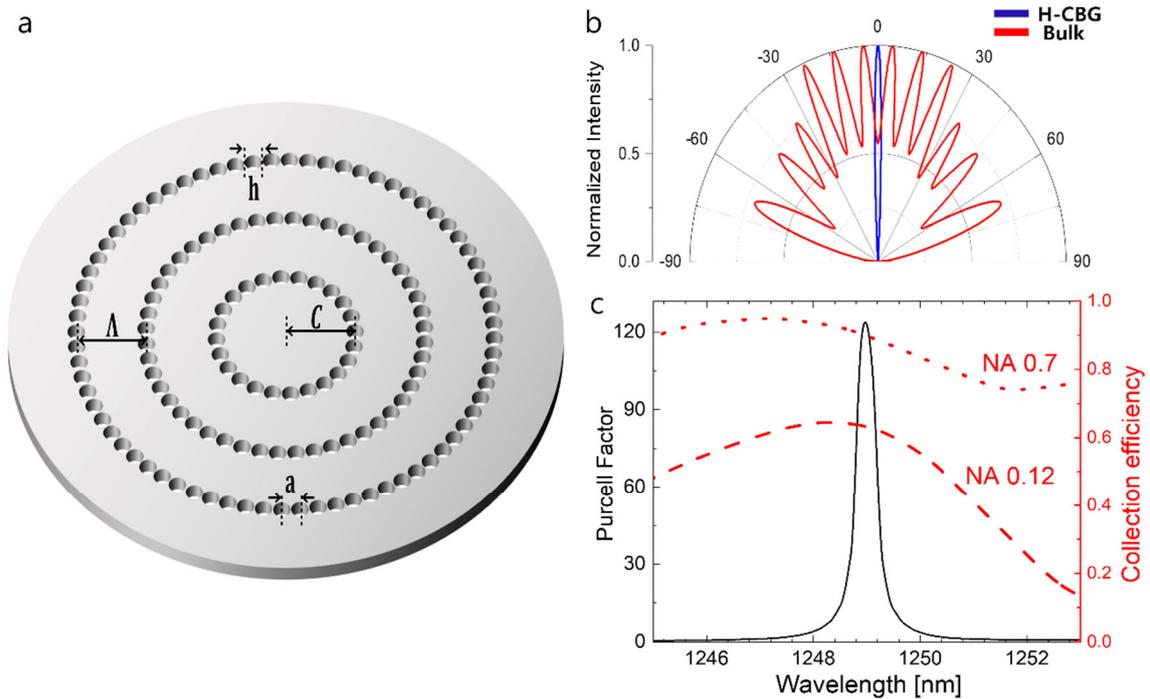

**Figure S4. Simulating parameters and numerical analysis of H-CBG** (a) Schematic of a H-CBG cavity with key parameters: radial period (**Λ**), center disk radius (***c***), hole size (***h***), and axial period (***a***). (b) Normalized angular distribution of the far-field emission from a dipole in an H-CBG cavity (blue) and a bulk sample (red). (c) Purcell factor (black solid line) and collection efficiency (red dashed/dotted lines) of the H-CBG as a function of wavelength for different numerical apertures (NA=0.12 and NA=0.7).

**(2) Optimization of a cavity mode for highly directional emission**

We simulated how the single-photon coupling efficiency decreases with the misalignment of a quantum dot and a cavity from a fiber core. In this simulation, the core diameter of the SMF-28 fiber was set to 8.14 μm, with refractive indices of the core and cladding set to 1.452 and 1.447, respectively (NA = 0.12). **Figure S5(a)** illustrates the misalignment of the quantum dot from the cavity's center and plots the fiber-coupling efficiency as a function of the quantum dot's position from the center. Additionally, we calculated the fiber-coupling efficiency with respect to the alignment mismatch between the center of the fiber core and the cavity, while the quantum dot is positioned at the center of the cavity (Figure S5 (b)).

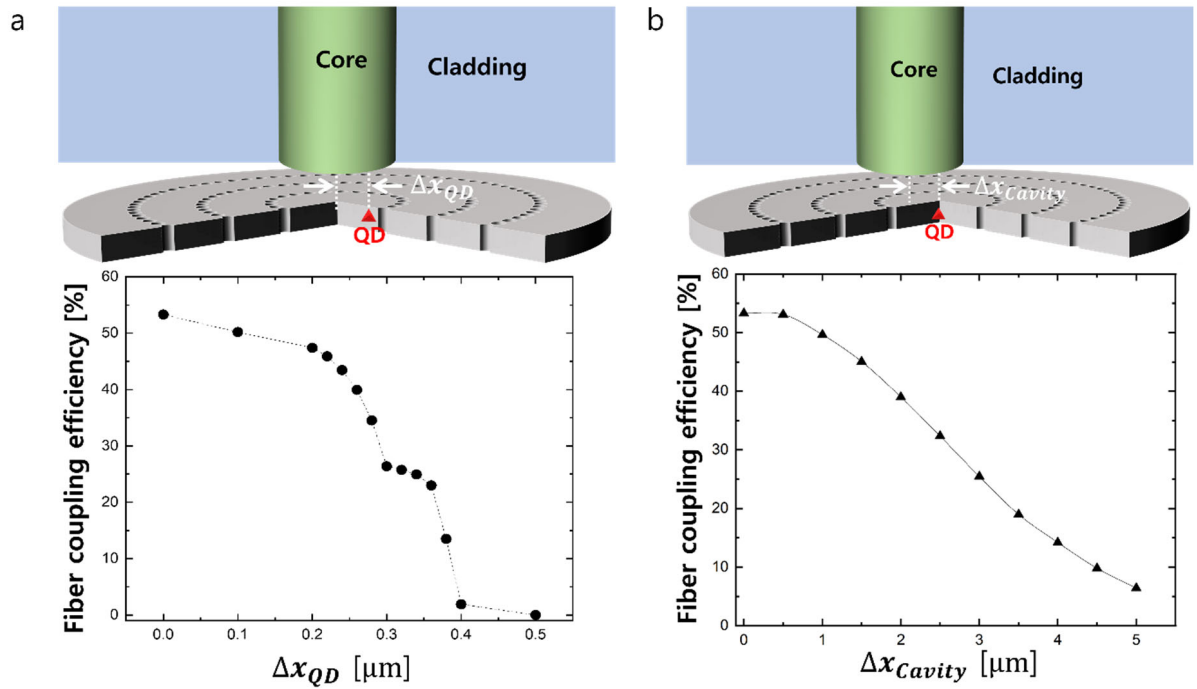

**Figure S5. Numerical analysis of coupling efficiency** (a) Fiber coupling efficiency as a function of the quantum dot position ($\Delta x_{QD}$) relative to the fiber core. (b) Fiber coupling efficiency as a function of cavity misalignment ($\Delta x_{cavity}$) relative to the fiber core.

## 4. Hong-Ou-Mandel (HOM) interferometer

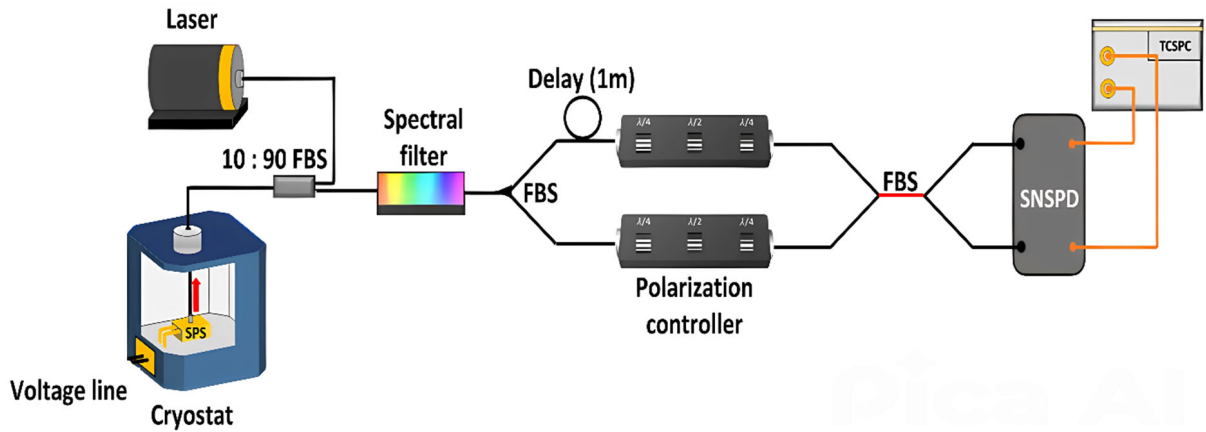

**Figure S6. A schematic of an asymmetric Mach-Zehnder type fiber-based HOM interferometer with a 1 m delay line.** FBS: Fiber beam splitter, TCSPC: time-correlated single-photon counter, SNSPD: superconducting nanowire single-photon detector